\documentclass[10pt,leqno]{article}

\usepackage[a4paper,
            left=22mm,
            right=22mm,
            top=25mm,
            bottom=25mm]{geometry}

\usepackage{graphicx}
\usepackage{cite}
\usepackage{amsmath,amssymb,amsthm}
\usepackage{booktabs}
\usepackage{tabularx}
\usepackage{adjustbox}
\usepackage{xcolor}
\usepackage{hyperref}
\usepackage{placeins}

\hypersetup{
    colorlinks=true,
    linkcolor=black,
    citecolor=black,
    urlcolor=black
}

\newcommand{\numberthis}{%
  \addtocounter{equation}{1}%
  \tag{\theequation}%
}

\begin{document}
\title{Global geometry of the genotype–phenotype map illuminates a trade-off between penetrance and mutational adaptability}

\author{%
\normalsize
Yutaro Ikeda$^{1,2,3}$, Kunihiko Kaneko$^{4}$, Tetsuhiro S. Hatakeyama$^{5}$\\
\footnotesize
\makebox[\textwidth][c]{%
\parbox{0.92\textwidth}{\centering
$^{1}$ Department of Physics, Graduate School of Science, The University of Tokyo, 7-3-1 Hongo, Bunkyo-ku, Tokyo 113-0033, Japan \\
$^{2}$ Universal Biology Institute, Graduate School of Science, The University of Tokyo, 7-3-1 Hongo, Bunkyo-ku, Tokyo 113-0033, Japan \\
$^{3}$ Center for Biosystems Dynamics Research, RIKEN, 2-2-3 Minatojima-minamimachi, Chuo-ku, Kobe, Hyogo 650-0047, Japan \\
$^{4}$ Niels Bohr Institute, University of Copenhagen, Jagtvej 128, 2200 Copenhagen, Denmark \\
$^{5}$ Earth-Life Science Institute, Institute of Future Science, Institute of Science Tokyo, 2-12-1-IE-1 Ookayama, Meguro-ku, Tokyo 152-8550, Japan \\
\texttt{ikeda@ubi.s.u-tokyo.ac.jp}
\texttt{kaneko@complex.c.u-tokyo.ac.jp}
\texttt{hatakeyama@elsi.jp}
}
}
}

\date{}

\maketitle

\begin{abstract}
Evolution in changing environments requires both reliable expression of the currently favored phenotype, as quantified by penetrance, and the capacity to reach alternative phenotypes through mutation. Previous studies suggest that high penetrance may restrict such mutational access. However, because these studies focus on evolved genotypes and local mutational neighborhoods, they cannot determine whether this local constraint limits mutational adaptability under environmental change. Such adaptability depends on a genotype's position relative to high-fitness regions associated with other environments. Addressing this question requires reconstructing the full probability distribution over phenotypes for every genotype and the resulting environment-specific fitness landscapes across genotype space. Such reconstruction is generally infeasible because genotype and phenotype spaces grow combinatorially. Here, an abstract model of stochastic genotype-phenotype mapping, inspired by interacting spins in statistical physics, permits exhaustive reconstruction of the map. We find that high-penetrance genotypes tend to occupy the interior of environment-specific high-fitness regions and are mutationally robust, whereas lower-penetrance genotypes tend to lie near their boundaries and have greater mutational access to high-fitness regions associated with alternative environments. This global geometry generates a trade-off between penetrance and mutational adaptability. In evolutionary simulations, stronger phenotypic noise in a fixed environment increases the selective advantage of reliable expression, thereby favoring high penetrance and mutational robustness. Frequent environmental change instead favors mutational accessibility at the expense of penetrance. Thus, penetrance and adaptability are opposing consequences of the same global geometry, and environmental conditions determine the evolutionary balance between them, thereby explaining when genotypes with incomplete penetrance can be favored.
\end{abstract}

\section*{Introduction}

Organisms are exposed to noise arising both from intrinsic biological processes and from changes in their environments. In a given environment, intrinsic noise in molecular and developmental processes, including thermal fluctuations and stochastic gene expression, can cause the same genotype to produce different phenotypes. Because such variation can reduce the probability of expressing an adaptive phenotype and thereby lower fitness, selection favors genotypes that reliably express the appropriate phenotype. When the environment changes, however, the adaptive phenotype may also change, and populations must retain the capacity to reach alternative phenotypes through mutation. In fact, phenotypic reliability has long been studied in genetics, where penetrance, the probability that a genotype expresses a particular phenotype, provides a quantitative measure \cite{Whitten1968-tk, Elowitz2002-ws, Eldar2009-mc, Raj2010-wu, Sucharov2019-bd}. The capacity to adapt to environmental change through mutation, which we refer to as mutational adaptability, has likewise long been regarded as central to evolutionary biology \cite{Earl2004-ah, Kumawat2025-up, Ciliberti2007-zm}. How these two properties are related remains unclear.

Pioneering theoretical work suggested that high penetrance may come at the cost of mutational adaptability. In a model of RNA folding, Ancel and Fontana showed that selection for sequences that consistently adopt the same structure despite thermal fluctuations also increases mutational robustness. Mutant sequences therefore tend to retain the same structure and fitness in the original environment \cite{Ancel2000-on}, making them less likely to generate alternative phenotypes through mutation when environmental change favors a different phenotype. Similar associations between high penetrance and high mutational robustness have been reported across diverse systems, from theoretical models of gene regulatory networks to empirical biological systems \cite{Kaneko2007-pa, Kaneko2011-hv, Kaneko2012-ue, Ciliberti2007-tp, Szollosi2009-ks, Butkovic2020-in, Kaneko2021-mk, Kaneko2022-bl, Kaneko2025-ss}, suggesting that this coupling is not specific to RNA and may be broadly shared across biological systems.

However, these studies have largely focused on genotypes evolved under selection for specific phenotypes and on their local mutational neighborhoods \cite{Ancel2000-on, Kaneko2007-pa, Kaneko2011-hv, Kaneko2012-ue, Ciliberti2007-tp}. Because evolution samples only a restricted subset of genotypes, the observed relationship may reflect the particular regions reached by evolution rather than an intrinsic organization of genotype space. Determining whether the local coupling between penetrance and mutational robustness gives rise to a global constraint on mutational adaptability therefore requires a comprehensive characterization of the genotype-phenotype map (GP map), which describes the full set of phenotypes that each genotype can produce. For stochastic phenotypic expression, a global analysis of penetrance requires the full probability distribution over all possible phenotypes for every genotype, rather than only the most probable phenotype. Because both genotype and phenotype spaces grow combinatorially with system size, the number of genotype-phenotype pairs rapidly becomes enormous, making such an exhaustive reconstruction generally infeasible.

Resolving this global relationship is central to understanding the evolution of incomplete penetrance. In a fixed environment, selection can favor genotypes that express the currently adaptive phenotype with high penetrance. Yet incomplete penetrance is well documented and has been implicated in facilitating phenotypic transitions during evolution \cite{Elowitz2002-ws, Eldar2009-mc, Raj2010-wu, Beye2013-zs}. This raises the question of when genotypes with incomplete penetrance can be favored rather than selected against. If lower penetrance is globally associated with greater mutational access to alternative adaptive phenotypes, environmental change could provide an offsetting advantage. Such an association would imply an evolutionary trade-off between penetrance and mutational adaptability. Moreover, because frequent environmental change is known to favor genotypes with high mutational adaptability \cite{Earl2004-ah, Sachdeva2020-ts, Kumawat2025-up}, the balance between penetrance and mutational adaptability in that trade-off may depend on the environmental timescale.

Here, we address two questions: (1) how penetrance is organized across genotype space and related to mutational adaptability, and (2) how evolution under fixed and changing environments selects among different regions of this global structure.

To address these questions, we build on previous spin-based studies of mutational robustness in deterministic and probabilistic GP maps \cite{Sakata2009-aq, Pham2023-iu, Mohanty2023-ue, Sappington2025-ob} and use a statistical-physics model of interacting spins that permits exhaustive reconstruction of a stochastic GP map. By enumerating all genotypes and calculating the complete phenotype distribution for each genotype, we characterize penetrance and fitness throughout genotype space, without restricting the analysis to genotypes reached by a particular evolutionary history. We uncover a global geometry in which high penetrance is correlated with mutational robustness, whereas lower penetrance corresponds to greater mutational access to high-fitness regions defined by alternative environments.

Next, we use evolutionary simulations to determine how environmental conditions favor different regions of this global structure. Under a fixed environment, stronger phenotypic noise favors genotypes with high penetrance and high mutational robustness, whereas frequent environmental change favors genotypes with lower penetrance and higher mutational accessibility. We further find that these qualitative relationships persist in a larger system. Together, these results show that the global organization of the GP map gives rise to a trade-off between penetrance and mutational adaptability, and that environmental conditions determine the evolutionary balance between them.

\section*{Results}

\subsection*{Abstract GP map model}

We introduce an abstract GP map model in which a genotype $\boldsymbol{J}$ stochastically generates a multivariate phenotype $\boldsymbol{S}$. We distinguish the full phenotype $\boldsymbol{S}$ from its fitness-relevant focal components $\boldsymbol{S}_{\mathrm{focal}}$. The environment specifies a target pattern only for $\boldsymbol{S}_{\mathrm{focal}}$, so that multiple distinct full phenotypes can be equally adaptive. Genotypic fitness is defined as the total probability of expressing this adaptive focal trait.

We consider a system of $N$ globally interacting Ising spins as an abstract model of a GP map \cite{Sakata2009-aq}. In this model, the spin configuration $\boldsymbol{S}=\left(S_{1},\ldots, S_{N}\right)$ and the spin interactions $\boldsymbol{J}=\{J_{ij}\}$ represent the phenotype and genotype, respectively. For simplicity, we assume that $\boldsymbol{J}$ is symmetric with $J_{ii}=0$. $S_{i}$ and $J_{ij}\left(i\neq j\right)$ take values in $\{+1, -1\}$. These variables admit several biological interpretations. For example, in terms of gene expression, phenotype $\boldsymbol{S}$ corresponds to the gene expression pattern, and genotype $\boldsymbol{J}$ corresponds to regulatory interactions between genes: $S_{i}=+1$ means that gene $i$ is expressed, and $J_{ij}=J_{ji}=+1$ means that the expression states of genes $i$ and $j$ tend to be aligned.

At each generation, the genotype $\boldsymbol{J}$ is fixed, and the probability for each phenotype, given by the spin configuration $\boldsymbol{S}$, is determined by the energy as follows.
\begin{align*}
p\left(\boldsymbol{S}\mid\boldsymbol{J}, T_{p}\right)&=\frac{1}{Z_{S}(\boldsymbol{J},T_p)}\exp\left(-\frac{1}{T_{p}}E\left(\boldsymbol{S}; \boldsymbol{J}\right)\right), \\
Z_{S}(\boldsymbol{J},T_p)&=\sum_{\boldsymbol{S}}\exp\left(-\frac{1}{T_{p}}E\left(\boldsymbol{S}; \boldsymbol{J}\right)\right), \numberthis  \label{eqn:eq_prob_S} \\
E\left(\boldsymbol{S}; \boldsymbol{J}\right)&=-\frac{1}{\sqrt{N}}\sum_{i<j}{J_{ij}S_{i}S_{j}},
\end{align*}
where the parameter $T_{p}$ represents the noise level in phenotypic expression and is fixed. Due to global spin-flip symmetry, a pair of phenotypes $\boldsymbol{S}$ and $-\boldsymbol{S}$ has the same $E\left(\boldsymbol{S}; \boldsymbol{J}\right)$. Consequently, the penetrance is bounded above by $\frac{1}{2}$ in this model. This symmetry-imposed upper limit is common to all genotypes.

Next, we introduce fitness, a key quantity in the evolutionary dynamics. We regard a phenotype that matches the desired pattern specified by the current environment as adapted to that environment. We define the phenotypic adaptation level $F\left(\boldsymbol{S}; \boldsymbol{\tau}\right)$ as a binary variable based on the match between $N_{\mathrm{focal}}$ elements of the phenotype $\boldsymbol{S}_{\mathrm{focal}} = \left(S_{1},\ldots,S_{N_{\mathrm{focal}}}\right)$ and a given target pattern $\boldsymbol{\tau} = \left(\tau_{1},\ldots,\tau_{N_{\mathrm{focal}}}\right) \in \{+1,-1\}^{N_{\mathrm{focal}}}$, which represents the desired phenotype in a given environment. $F\left(\boldsymbol{S}; \boldsymbol{\tau}\right)$ is defined as
\begin{align*}
F\left(\boldsymbol{S}; \boldsymbol{\tau}\right)=\delta\left(\boldsymbol{S}_{\mathrm{focal}}, \boldsymbol{\tau}\right)+\delta\left(\boldsymbol{S}_{\mathrm{focal}}, -\boldsymbol{\tau}\right), \numberthis \label{eqn:eq_fitness_S}
\end{align*}
where $\delta\left(\boldsymbol{x},\boldsymbol{x}'\right)=1$ if $\boldsymbol{x}=\boldsymbol{x}'$ and $0$ otherwise. Since there is global spin-flip symmetry, we treat $\boldsymbol{\tau}$ and $-\boldsymbol{\tau}$ as equivalent target patterns: i.e., $\boldsymbol{S}$ and $-\boldsymbol{S}$ are realized with equal probability. Thus, there are $2^{N_{\mathrm{focal}}-1}$ possible target patterns.

Since each phenotype is expressed with genotype-dependent probability in this model, the genotypic fitness $\langle{F}\rangle\left(\boldsymbol{J}; \boldsymbol{\tau}, T_{p}\right)$, which governs the evolutionary dynamics, is defined as the expected phenotypic adaptation level over the probability distribution of phenotypes. That is, 
\begin{align*}
\langle{F}\rangle\left(\boldsymbol{J}; \boldsymbol{\tau}, T_{p}\right)
&:=\sum_{\boldsymbol{S}}F\left(\boldsymbol{S}; \boldsymbol{\tau}\right)p\left(\boldsymbol{S}\mid\boldsymbol{J}, T_{p}\right) \numberthis \label{eqn:eq_fitness_J} \\
&=P\left(\boldsymbol{S}_{\mathrm{focal}}=\pm\boldsymbol{\tau}\mid\boldsymbol{J},T_{p}\right).
\end{align*}
Eq.~\ref{eqn:eq_fitness_J} therefore defines fitness as the total probability mass assigned to these adaptive phenotypes. In contrast, penetrance, defined below, refers to the probability of a particular full phenotype. This distinction is deliberate: the environment directly evaluates only the focal components of a multidimensional phenotype, whereas the reproducibility of the full phenotype is not directly selected. The non-focal components are not directly scored by the fitness function, although they can influence fitness through the genotype-dependent distribution $p\left(\boldsymbol{S}\mid\boldsymbol{J}, T_{p}\right)$.

Genotypic mutation, which flips an element $J_{ij}$ in $\boldsymbol{J}$ and its symmetric counterpart $J_{ji}$, alters the probability distribution of phenotypes and thereby can change genotypic fitness.

\begin{table}[tbp]
\centering
\caption{Elements in the abstract GP map model}
\begin{tabularx}{\textwidth}{@{}l X X@{}}
\toprule
\multicolumn{1}{c}{Symbol} &  \multicolumn{1}{c}{Description} & \multicolumn{1}{c}{} \\
\midrule
$\boldsymbol{S}$ & Phenotype & $\boldsymbol{S}:=\left(S_{1},\ldots, S_{N}\right), S_{i}\in\{-1, +1\}$ \\
$E\left(\boldsymbol{S};\boldsymbol{J}\right)$ & Energy of phenotype $\boldsymbol{S}$ on genotype $\boldsymbol{J}$ & $E\left(\boldsymbol{S};\boldsymbol{J}\right)\in \mathbb{R}$ \\
$p\left(\boldsymbol{S}\mid\boldsymbol{J}, T_{p}\right)$ & Probability that genotype $\boldsymbol{J}$ expresses phenotype $\boldsymbol{S}$ & $0 \leq p\left(\boldsymbol{S}\mid\boldsymbol{J}, T_{p}\right) \leq \frac{1}{2}$ \\
$\boldsymbol{\tau}$ & Target pattern & $\boldsymbol{\tau}:=\left(\tau_{1},\ldots, \tau_{N_{\mathrm{focal}}}\right), \tau_{i}\in\{-1, +1\}$  \\
$F\left(\boldsymbol{S}; \boldsymbol{\tau}\right)$ & Adaptation level of phenotype $\boldsymbol{S}$ & $F\left(\boldsymbol{S}; \boldsymbol{\tau}\right)\in\{0, 1\}$ \\
$\langle{F}\rangle\left(\boldsymbol{J}; \boldsymbol{\tau}, T_{p}\right)$ & Fitness of genotype $\boldsymbol{J}$ (expectation of $F\left(\boldsymbol{S}; \boldsymbol{\tau}\right)$ over phenotype $\boldsymbol{S}$ on $\boldsymbol{J}$) & $0 \leq \langle{F}\rangle\left(\boldsymbol{J}; \boldsymbol{\tau}, T_{p}\right) \leq 1$ \\
$w\left(\boldsymbol{J}\rightarrow\boldsymbol{J}'\right)$ & Transition probability from $\boldsymbol{J}$ to $\boldsymbol{J}'$ & $0 \leq w\left(\boldsymbol{J}\rightarrow\boldsymbol{J}'\right) \leq 1$ \\
$\boldsymbol{J}$ & Genotype & $\boldsymbol{J}:=\{J_{ij}\}, J_{ii}=0, J_{ij}\in\{-1, +1\} \quad (i \neq j)$ \\
\midrule
$N$ & Number of spins &  \\
$N_{\mathrm{focal}}$ & Number of focal spins &  \\
$T_{p}$ & Phenotypic noise &  \\
$T_{g}$ & Parameter controlling selection strength &  \\
$t_{\mathrm{interval}}$ & Interval of periodic environmental switching &  \\
\bottomrule
\end{tabularx}
\end{table}

\subsection*{Evaluation of penetrance from the phenotypic landscape}

In this paper, we define penetrance as the probability that genotype $\boldsymbol{J}$ expresses phenotype $\boldsymbol{S}$, $p\left(\boldsymbol{S}\mid\boldsymbol{J}, T_{p}\right)$. High penetrance means that a phenotype is expressed with high probability even under phenotypic noise. Conversely, when the penetrance of a phenotype is low, other phenotypes can be expressed with high probability. Hence, it is necessary to measure its probability relative to those of alternative phenotypes.

In this model, the ratio of the probabilities of phenotypes $\boldsymbol{S}$ and $\boldsymbol{S}'$ for a given genotype $\boldsymbol{J}$ is proportional to $\exp\left(-\frac{1}{T_{p}}\left(E\left(\boldsymbol{S}; \boldsymbol{J}\right)-E\left(\boldsymbol{S}'; \boldsymbol{J}\right)\right)\right)$, as follows from Eq.~\ref{eqn:eq_prob_S}. We focus on the phenotypes expressed with the highest probability by a given genotype, which correspond to the minimum-energy states \footnote{Due to global spin-flip symmetry, minimum-energy states occur in spin-reversed pairs.}. We hereafter refer to these phenotypes as the dominant phenotypes and investigate the penetrance of a dominant phenotype for each genotype.

To examine the penetrance of each phenotype for a given genotype, we calculate the energy of each phenotype and represent these energies as a landscape over phenotypes $\boldsymbol{S}$ of a given $\boldsymbol{J}$ in this model. We refer to this energy landscape over phenotype space under a fixed genotype as the "phenotypic landscape". The phenotypic landscape allows us to quantify the difference in probability of each phenotype and the penetrance of a dominant phenotype. 

To reveal how penetrance evolves, we comprehensively examine the phenotypic and fitness landscapes across the entire genotype space. To enumerate both landscapes over all phenotypes $\boldsymbol{S}$ and genotypes $\boldsymbol{J}$, we use a small system with $N=6, N_{\mathrm{focal}}=3$. To focus on the effect of the phenotypic landscape on penetrance, we initially set the phenotypic noise to a low value, $T_{p}=0.03125$.

\begin{figure}[htp]
\centering
\includegraphics[width=.70\linewidth]{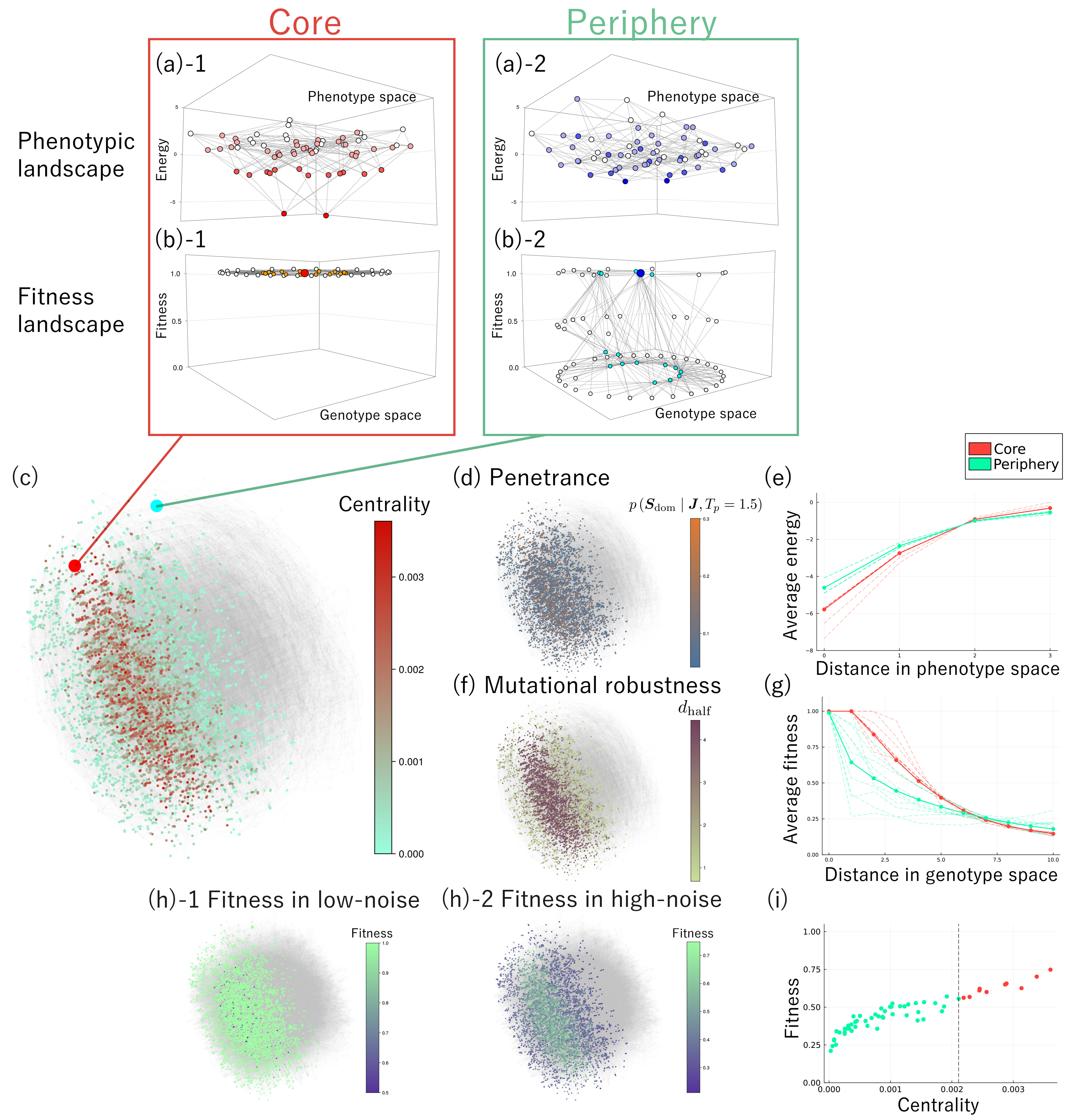}
\caption{(a) Phenotypic landscape of representative genotypes $\boldsymbol{J}_{\mathrm{f}}$((a)-1) , $\boldsymbol{J}_{\mathrm{r}}$((a)-2). For a comprehensive analysis across genotype space, we set $N=6$ and $N_{\mathrm{focal}}=3$. Nodes represent phenotypes $\boldsymbol{S}$, and edges connect pairs of phenotypes that differ only in one element $S_{i}$. The height and node color represent energy $E\left(\boldsymbol{S}; \boldsymbol{J}\right)$ and phenotypic adaptation level $F\left(\boldsymbol{S}; \boldsymbol{\tau}_{1}\right)$, respectively. Because a pair of phenotypes $\boldsymbol{S}$ and $-\boldsymbol{S}$ has the same $E\left(\boldsymbol{S}; \boldsymbol{J}\right)$ due to global spin-flip symmetry, there are an even number of minima in a phenotypic landscape. (b) Local fitness landscapes around representative genotypes $\boldsymbol{J}_{\mathrm{f}}$((b)-1) and $\boldsymbol{J}_{\mathrm{r}}$((b)-2) up to $d=2$. To calculate fitness, we use $\boldsymbol{\tau}_{1}=\left(+1, +1, +1\right)$. Due to gauge symmetry, the results of other $\boldsymbol{\tau}$ are the same as those of $\boldsymbol{\tau}_{1}$. Nodes represent genotypes and edges connect pairs of genotypes that differ only by interaction elements $J_{ij}$ and $J_{ji}$. The height and node color represent the genotypic fitness of a mutant $\langle{F}\rangle\left(\boldsymbol{J}_{\mathrm{mut}}; \boldsymbol{\tau}, T_{p}\right)$ at $T_{p}=0.03125$ and mutational distance, $d$, respectively. (c) Distribution of betweenness centrality in $G\left(\boldsymbol{\tau}_{1}\right)$ on genotype space. To connect this analysis to our later discussion of adaptability to the alternative environment, we display $G\left(\boldsymbol{\tau}_{1}\right)$ as part of $G\left(\boldsymbol{\tau}_{1},\boldsymbol{\tau}_{2}\right)$, the network obtained by excluding high-fitness regions associated with target patterns other than $\boldsymbol{\tau}_{1}$ and $\boldsymbol{\tau}_{2}$, where $\boldsymbol{\tau}_{2}=\left(+1,+1,-1\right)$ (see Materials and Methods). $\boldsymbol{J}_{\mathrm{f}}$ and $\boldsymbol{J}_{\mathrm{r}}$ in (a) and (b) are highlighted by large cyan and orange points. (d) Distribution of penetrance of a dominant phenotype $\boldsymbol{S}_{\mathrm{dom}}$ in $G\left(\boldsymbol{\tau}_{1}\right)$ on genotype space in the same manner as (c). Genotype centrality and penetrance of a dominant phenotype are correlated in genotype space. (e) Average energy of phenotypes at each minimum Hamming distance to any dominant phenotype $l$. The dashed lines are examples of individual genotypes, and the solid lines are averages over core genotypes and peripheral genotypes. (f) Distribution of half-fitness mutational distance $d_{\mathrm{half}}$ in $G\left(\boldsymbol{\tau}_{1}\right)$ on genotype space in the same manner as (c). Genotype centrality and half-fitness mutational distance are correlated in genotype space. (g) Average mutant fitness for each $d$. The dashed lines show examples of individual genotypes, and solid lines show averages over core genotypes and peripheral genotypes. (h) Distribution of fitness in environment $\boldsymbol{\tau}_{1}$ on $G\left(\boldsymbol{\tau}_{1},\boldsymbol{\tau}_{2}\right)$ in low-noise ($T_{p}=0.03125$, (h)-1) and high-noise ($T_{p}=1.5$, (h)-2) conditions, represented by node color. (i) The relationship between centrality and fitness in high-noise conditions.}
\label{fig:Fixed}
\end{figure}

\subsection*{Genotypes exhibit distinct phenotypic landscapes and penetrance}

First, to determine whether the penetrance depends on the genotype, we calculate the energies $E\left(\boldsymbol{S}; \boldsymbol{J}\right)$ of all possible phenotypes for each genotype, which gives the phenotypic landscape. Phenotypic landscapes vary in the energy separation between dominant phenotypes and alternative phenotypes. To make this variation concrete, we first examine two representative genotypes with maximum and minimum genotype fitness under the high-noise condition ($T_{p}=1.5$), $\boldsymbol{J}_{\mathrm{f}}$ and $\boldsymbol{J}_{\mathrm{r}}$, respectively. These examples serve to motivate the global analysis below; the general relationship is tested over all genotypes in the subsequent sections.

The phenotypic landscape of $\boldsymbol{J}_{\mathrm{f}}$ has large energy gaps between the dominant phenotypes and the other phenotypes (Fig.~\ref{fig:Fixed}a-1). In addition, in this phenotypic landscape, phenotypes farther from the dominant phenotype $\boldsymbol{S}_\mathrm{dom}$ have higher energy. We call this type of phenotypic landscape a "funnel-like" phenotypic landscape, in which the penetrance of the dominant phenotypes is high.

On the other hand, the phenotypic landscape of $\boldsymbol{J}_{\mathrm{r}}$ has smaller energy gaps between the dominant phenotypes and the other phenotypes than that of $\boldsymbol{J}_{\mathrm{f}}$ (Fig.~\ref{fig:Fixed}a-2). Also, the energy of a phenotype does not increase monotonically with its minimum Hamming distance from the dominant phenotypes $\boldsymbol{S}_{\mathrm{dom}}$ in this phenotypic landscape. We call this type of phenotypic landscape a "rugged" phenotypic landscape, in which the penetrance of the dominant phenotypes is low.

These results suggest that penetrance depends on genotype and is an evolvable property.

\subsection*{Mutational robustness is associated with penetrance}

Next, we consider which genotypes exhibit more funnel-like or rugged phenotypic landscapes. A more funnel-like phenotypic landscape is characterized not only by high penetrance of the dominant phenotype, but also by high phenotypic robustness: the dominant phenotype remains highly probable under increased noise in expression. Previous studies have suggested that phenotypic robustness is correlated with and evolves together with the robustness of genotypic fitness to mutation, i.e., mutational robustness \cite{Waddington1942-cz, Ancel2000-on, Kaneko2007-pa, Szollosi2009-ks, Kaneko2011-hv, Kaneko2012-ue, Butkovic2020-in, Kaneko2021-mk, Kaneko2022-bl, Ciliberti2007-tp, Kaneko2025-ss}. Thus, the mutational robustness of genotypes may indicate the type of their phenotypic landscape and their penetrance. Using this model, we can confirm this relationship between phenotypic landscape and mutational robustness by comprehensively quantifying both across the entire genotype space. Mutational robustness can be quantified by computing the fitness landscape.

To verify the relationship between penetrance and mutational robustness, we calculate the fitness of all mutants $\langle{F}\rangle\left(\boldsymbol{J}_{\mathrm{mut}}; \boldsymbol{\tau}_{1}, T_{p}\right)$ and average that fitness over all mutants at each mutational distance $d$, where the mutant genotype $\boldsymbol{J}_{\mathrm{mut}}$ differs from the original genotype $\boldsymbol{J}$ by $d$ independent elements. Fig.~\ref{fig:Fixed}b shows fitness landscapes up to $d=2$ for each of the genotypes, exhibiting funnel-like (high-penetrance) and rugged (low-penetrance) phenotypic landscapes, $\boldsymbol{J}_{\mathrm{f}}$ and $\boldsymbol{J}_{\mathrm{r}}$, as examples.

The fitness landscape around $\boldsymbol{J}_{\mathrm{f}}$ forms a high-fitness plateau, with fitness remaining nearly unchanged under a small number of mutations (Fig.~\ref{fig:Fixed}b-1). We refer to this type of fitness landscape as a "plateau-like" fitness landscape. This fitness landscape indicates that $\boldsymbol{J}_{\mathrm{f}}$ has high mutational robustness.

On the other hand, there are multiple local maxima, and fitness decreases sharply within a few mutations of $\boldsymbol{J}_{\mathrm{r}}$ in the fitness landscape around $\boldsymbol{J}_{\mathrm{r}}$ (Fig.~\ref{fig:Fixed}b-2). In this fitness landscape, many mutants at $d=1$ show reduced fitness but partially recover at $d=2$. We refer to this type of fitness landscape as a "rugged" fitness landscape. This fitness landscape indicates that $\boldsymbol{J}_{\mathrm{r}}$ has low mutational robustness.

These results suggest that mutational robustness correlates with penetrance. Genotypes with higher penetrance have higher mutational robustness. 

\FloatBarrier

\subsection*{Penetrance and mutational robustness correlate with position in genotype space}

To investigate the generality of the correlation between penetrance and mutational robustness across genotype space, we examine how each type of genotype is distributed. Since mutations correspond to movements in genotype space, the distribution in genotype space may be associated with mutational robustness.

The observation in Fig.~\ref{fig:Fixed}b-1 suggests that mutational neighbors of a genotype with high penetrance tend to have high fitness. If so, such a genotype is surrounded by genotypes with high fitness. Thus, we hypothesize that genotypes with high penetrance are positioned at the center of the high-fitness region.

To test this hypothesis, we characterize each genotype using centrality in the high-fitness region under a given environment. We investigate the relationship between such centrality and the shapes of the phenotypic and fitness landscapes representing penetrance and mutational robustness. This analysis allows us to confirm how penetrance and mutational robustness are distributed and whether penetrance is associated with mutational robustness across genotype space.

We define the adaptive region under the low-noise condition, in which fitness is determined primarily by whether the dominant phenotypes match the target. This allows us to identify the genotypes adapted to $\boldsymbol{\tau}_{1}$ independently of the penetrance differences revealed under stronger phenotypic noise. We construct the genotype network $G\left(\boldsymbol{\tau}_{1}\right)$, whose nodes are genotypes satisfying $\langle{F}\rangle\left(\boldsymbol{J}; \boldsymbol{\tau}_1, T_{p}=0.03125\right)>0.5$, and whose edges connect pairs of genotypes that differ by a single mutation, i.e., by flipping one pair of symmetric interaction elements $J_{ij}$ and $J_{ji}$ (see Materials and Methods for its definition) \footnote{For all possible target patterns $\boldsymbol{\tau}_{i}$, $G\left(\boldsymbol{\tau}_{i}\right)$ is connected.}. 

To characterize the position of each genotype, we use betweenness centrality in $G\left(\boldsymbol{\tau}_{1}\right)$, which measures how frequently a node is included in the shortest paths between other nodes in the network \footnote{The betweenness centrality is $C_B\left(i\right)=\frac{2}{(n-1)(n-2)}\sum_{\substack{j<k\\ j,k\neq i}}\frac{\sigma_{jk}\left(i\right)}{\sigma_{jk}}$, where $\sigma_{jk}$ is the number of shortest paths from node $j$ to node $k$ and $\sigma_{jk}\left(i\right)$ is the number of shortest paths from node $j$ to node $k$ that pass through node $i$. $\frac{2}{(n-1)(n-2)}$ is a term for normalization.}. We define genotypes in the top $10\%$ of betweenness centrality within $G\left(\boldsymbol{\tau}_{1}\right)$ as "core genotypes", and the others as "peripheral genotypes". The representative genotypes discussed above, $\boldsymbol{J}_{\mathrm{f}}$ and $\boldsymbol{J}_{\mathrm{r}}$, belong to core and peripheral genotypes, respectively. Fig.~\ref{fig:Fixed}c shows the distribution of betweenness centrality in $G\left(\boldsymbol{\tau}_{1}\right)$. Fig.~\ref{fig:Fixed}c indicates that core genotypes occupy the interior of the high-fitness region, whereas peripheral genotypes tend to lie near its boundary. We can examine the relationship between centrality and penetrance (and mutational robustness) by comparing core and peripheral genotypes.

First, we test how penetrance depends on centrality. We calculate the penetrance of a dominant phenotype under the high-noise condition $T_{p}=1.5$, where the difference in the genotype-dependent phenotypic landscape is reflected in the penetrance. If the phenotypic landscape has a funnel-like shape, a dominant phenotype is faithfully expressed even under the high-noise condition. Fig.~\ref{fig:Fixed}d shows the relationship between genotype centrality and penetrance of one of the dominant phenotypes (see also SI Appendix, Fig. S1). In addition, we average the energy over phenotypes at each minimum Hamming distance to any dominant phenotype $l$, as shown in Fig.~\ref{fig:Fixed}e. These results indicate that genotypes with higher centrality in $G\left(\boldsymbol{\tau}_{1}\right)$ have higher penetrance of a dominant phenotype. 

We next examine how mutational robustness also depends on centrality. We average mutant fitness $\langle{F}\rangle\left(\boldsymbol{J}_{\mathrm{mut}}; \boldsymbol{\tau}_{1}, T_{p}=0.03125\right)$ over all mutants at each mutational distance $d$ from the original genotype. To quantify the mutational robustness, we define half-fitness mutational distance, $d_{\mathrm{half}}$, as the mutational distance at which the average mutant fitness decreases to half of the original genotypic fitness (see Materials and Methods). This metric captures how rapidly fitness decays away from the focal genotype in genotype space. Fig.~\ref{fig:Fixed}f shows the relationship between centrality and half-fitness mutational distance $d_{\mathrm{half}}$ (see also SI Appendix, Fig. S2). In addition, Fig.~\ref{fig:Fixed}g shows the relationship between $d$ and average mutant fitness at each $d$. These results indicate that genotypes with higher centrality in $G\left(\boldsymbol{\tau}_{1}\right)$ have more plateau-like fitness landscapes and higher mutational robustness. 

These results are consistent with the trend observed in Fig.~\ref{fig:Fixed}a and b, and indicate that the shapes of the phenotypic and fitness landscapes are correlated across the entire genotype space. This means that penetrance and mutational robustness change together. Importantly, this relationship is not imposed by the definition of fitness. Fitness is summed over all full phenotypes that express the target focal pattern, whereas penetrance concerns only the dominant full phenotype. A genotype can therefore, in principle, have high fitness while distributing its probability over several adaptive full phenotypes and consequently having low dominant-phenotype penetrance. The observed alignment between high centrality, high dominant-phenotype penetrance, and high mutational robustness is therefore an emergent property of the GP map.

\subsection*{High penetrance is selectively favored under a fixed environment}

To investigate evolutionary changes of penetrance, we next analyze how genotypic fitness is distributed in genotype space.

In this model, while the phenotypic landscape representing the energy of each phenotype is independent of $T_{p}$, genotypic fitness $\langle{F}\rangle\left(\boldsymbol{J}; \boldsymbol{\tau}, T_{p}\right)$ depends on $T_{p}$, because the probability that a genotype expresses a given phenotype depends on $T_{p}$. A larger value of $T_p$ broadens the distribution over phenotypes and therefore corresponds to stronger phenotypic noise. We refer to $T_{p}=0.03125$ and $T_{p}=1.5$ as the low-noise and high-noise conditions, respectively, and compare genotypic fitness between these two conditions to examine the effect of $T_{p}$ on genotypic fitness. 

Fig.~\ref{fig:Fixed}h shows how fitness is distributed over the network $G\left(\boldsymbol{\tau}_{1}\right)$ under low- and high-noise conditions, and Fig.~\ref{fig:Fixed}i shows the relationship between centrality and fitness under the high-noise condition (see also SI Appendix, Fig. S3). 

Under the low-noise condition, where the difference in probability between phenotypes induced by the energy gap is large, genotypes have almost the same fitness regardless of centrality (Fig.~\ref{fig:Fixed}h-1). Consequently, the probability of dominant phenotypes collectively is almost $1$. In that case, from Eq.~\ref{eqn:eq_fitness_J}, fitness $\langle{F}\rangle\left(\boldsymbol{J}; \boldsymbol{\tau}, T_{p}\right)$ is mostly determined by whether the dominant phenotypes match the target pattern $\boldsymbol{\tau}_{1}$ and takes a value $\langle{F}\rangle\left(\boldsymbol{J}; \boldsymbol{\tau}, T_{p}\right)\approx0$ or $\langle{F}\rangle\left(\boldsymbol{J}; \boldsymbol{\tau}, T_{p}\right)\approx1$. These results suggest that fitness is less sensitive to the shape of the phenotypic landscape under the low-noise condition.

In contrast, under the high-noise condition, centrality is correlated with fitness, and the fitness of core genotypes in $G\left(\boldsymbol{\tau}_{1}\right)$ is high, whereas $\langle{F}\rangle\left(\boldsymbol{J}; \boldsymbol{\tau}_{1}, T_{p}\right) \approx 0.5$ for peripheral genotypes in $G\left(\boldsymbol{\tau}_{1}\right)$ (Fig.~\ref{fig:Fixed}h-2, Fig.~\ref{fig:Fixed}i). In this condition, the differences in probability between phenotypes are smaller, so fitness $\langle{F}\rangle\left(\boldsymbol{J}; \boldsymbol{\tau}, T_{p}\right)$ also depends on phenotypes other than the dominant phenotypes. Thus, only core genotypes with larger energy gaps separating the dominant phenotypes from other phenotypes, which correspond to a funnel-like phenotypic landscape, can maintain high penetrance of the dominant phenotypes adapted to the environment and therefore have high fitness.

\begin{figure}[htp]
\centering
\includegraphics[width=.30\linewidth]{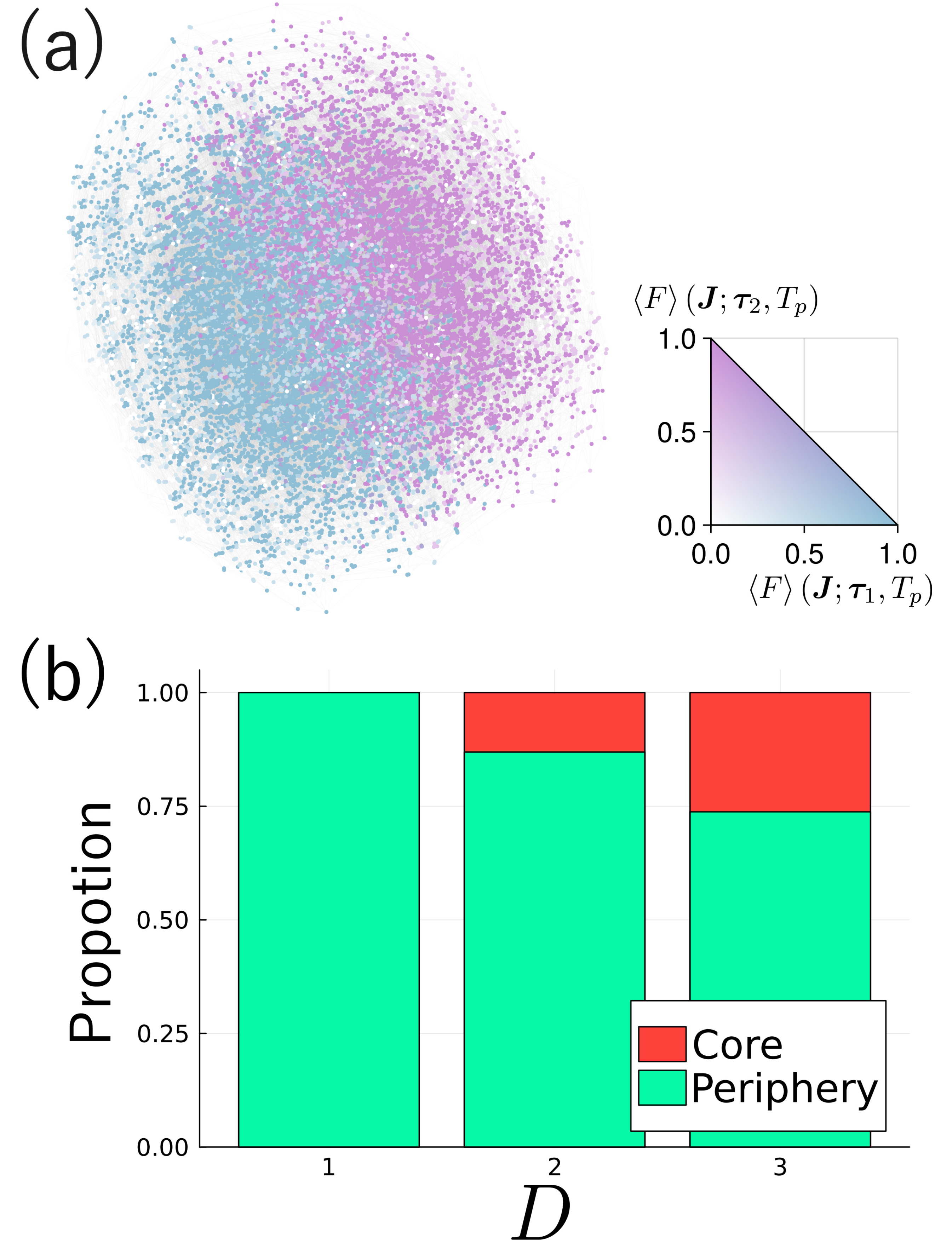}
\caption{(a) Fitness in $\boldsymbol{\tau}_{1}$ and $\boldsymbol{\tau}_{2}$ on the network $G\left(\boldsymbol{\tau}_{1},\boldsymbol{\tau}_{2}\right)$ in the same manner as Fig.~\ref{fig:Fixed}c. (b) Fraction of core and peripheral genotypes with each $D$ in all genotypes in $G\left(\boldsymbol{\tau}_{1}\right)$.}
\label{fig:Adaptability}
\end{figure}

\FloatBarrier

\subsection*{The penetrance map predicts access to adaptive regions under an alternative environment}

The above results indicate that core genotypes, whose dominant phenotypes have high penetrance and are adaptive, are favored in a fixed environment, especially under high phenotypic noise. In contrast, evolving organisms in nature often experience environmental change, such as climate change or the emergence of predators, that can alter which phenotype is adaptive. Therefore, genotypes with very high penetrance may be disadvantageous in terms of mutational adaptability in changing environments. 

To investigate mutational adaptability in changing environments, we first focus on the adaptation process to the new target pattern $\boldsymbol{\tau}_{2}=\left(+1, +1, -1\right)$ from genotypes with high fitness under a given target pattern $\boldsymbol{\tau}_{1}=\left(+1, +1, +1\right)$, and we examine how the centrality of a genotype in network $G\left(\boldsymbol{\tau}_{1}\right)$ influences the adaptation process.

Fig.~\ref{fig:Adaptability}a shows that $G\left(\boldsymbol{\tau}_{1}\right)$ and $G\left(\boldsymbol{\tau}_{2}\right)$ form distinct clusters in $G\left(\boldsymbol{\tau}_{1}, \boldsymbol{\tau}_{2}\right)$ which contains the high-fitness regions $G\left(\boldsymbol{\tau}_{1}\right)$ and $G\left(\boldsymbol{\tau}_{2}\right)$. These clusters are connected at the boundary between them. 

If a genotype in $G\left(\boldsymbol{\tau}_1\right)$ is close to the boundary between these clusters in $G\left(\boldsymbol{\tau}_{1}, \boldsymbol{\tau}_{2}\right)$, that genotype is closer to $G\left(\boldsymbol{\tau}_{2}\right)$ in genotype space and is expected to be more mutationally accessible to $\boldsymbol{\tau}_2$. Together with the results in Fig.~\ref{fig:Fixed}, such a genotype is expected to be a peripheral genotype. We therefore define the minimum Hamming distance in the full genotype space to any genotype in $G\left(\boldsymbol{\tau}_{2}\right)$ as a geometrical measure related to mutational adaptability, $D$. Because mutations change one interaction at a time, $D$ gives the minimum number of mutational steps required to enter the high-fitness region $G\left(\boldsymbol{\tau}_{2}\right)$. We therefore use $D$ as a geometric proxy for mutational accessibility, rather than as a complete measure of the adaptation rate. To test the expectation described above, we calculate $D$ for each genotype in $G\left(\boldsymbol{\tau}_{1}\right)$ (see also SI Appendix, Fig. S4). Fig.~\ref{fig:Adaptability}b shows the fractions of core and peripheral genotypes for each $D$ among all genotypes in $G\left(\boldsymbol{\tau}_{1}\right)$. Genotypes with smaller $D$, whose mutational accessibility to the new target pattern $\boldsymbol{\tau}_2$ is expected to be high, tend to be peripheral genotypes in $G\left(\boldsymbol{\tau}_{1}\right)$. 

These results suggest that peripheral genotypes with higher mutational accessibility to environmental change tend to have lower penetrance. Here, boundary proximity alone is not the central result. Rather, the spatial alignment between low penetrance and proximity to an alternative adaptive region can cause a trade-off between reliable phenotypic expression and mutational adaptability to other environments.

\begin{figure}[htp]
\centering
\includegraphics[width=.95\linewidth]{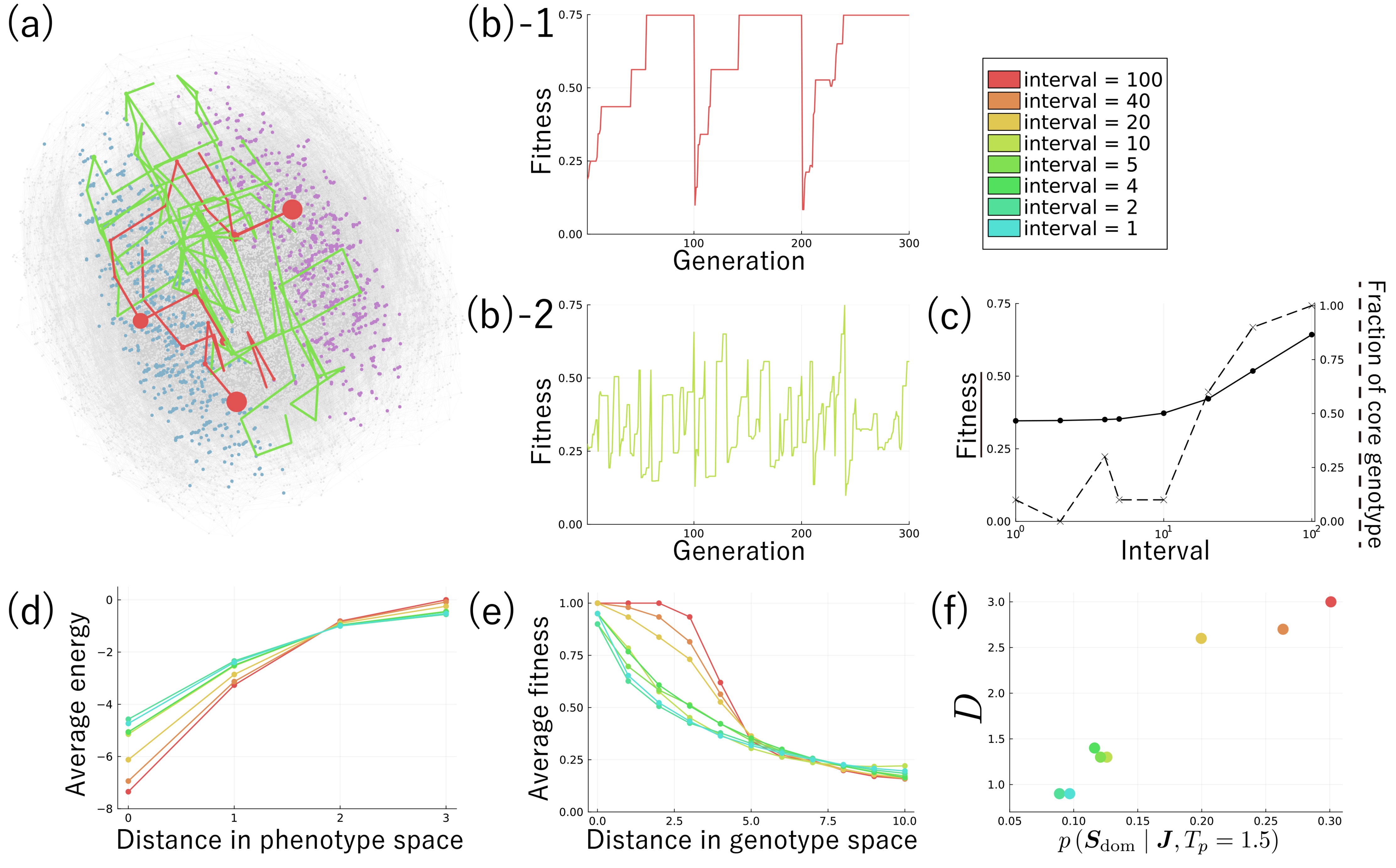}
\caption{(a) Representative trajectories of evolution under periodic environmental switching for $t_{\mathrm{interval}}=100$(red) and $10$(green) on the network $G\left(\boldsymbol{\tau}_{1},\boldsymbol{\tau}_{2}\right)$ in the same manner as Fig.~\ref{fig:Fixed}c. The size of nodes passed through is proportional to the number of generations spent. (b) Representative fitness trajectories of evolution under periodic environmental switching for $t_{\mathrm{interval}}=100$((b)-1) and $10$((b)-2). (c) Genotypic fitness in the current environment averaged over $1000$ evolutionary trajectories for each $t_{\mathrm{interval}}$ (solid line). Fraction of trajectories ending at a core genotype in $G\left(\boldsymbol{\tau}_{\mathrm{preferred}}\left(\boldsymbol{J}\right)\right)$ in $1000$ evolutionary trajectories (dashed line). (d) Average energy profile of the phenotypic landscape over genotypes obtained from $1000$ evolutionary trajectories after $10^4$ generations for each $t_{\mathrm{interval}}$ in the same manner as Fig.~\ref{fig:Fixed}e. (e) Average local fitness profile over genotypes obtained from $1000$ evolutionary trajectories after $10^4$ generations for each $t_{\mathrm{interval}}$ in the same manner as Fig.~\ref{fig:Fixed}g. We compute the local fitness landscape in the environment where each genotype has the highest fitness after $10^4$ generations.  (f) Relationship between the mean penetrance of a dominant phenotype and mean $D$ over genotypes evolved in periodic environmental switching after $10^4$ generations for each $t_{\mathrm{interval}}$.}
\label{fig:Repeating}
\end{figure}

\subsection*{How penetrance evolves depends on the timescale of environmental change}

In the analyses described above, we investigate the distribution of various quantities, such as penetrance and mutational robustness, in genotype space. Now, we examine how penetrance changes over generations through evolutionary simulations under environmental changes (see Materials and Methods).

Genotypes whose dominant phenotypes have high penetrance and are adaptive are likely to be selected under a fixed environment, as suggested in Fig.~\ref{fig:Fixed}, but evolution under frequent environmental changes may favor genotypes with low penetrance, which maintain mutational accessibility to other environments, as suggested in Fig.~\ref{fig:Adaptability}.

To verify this, we focus on evolution under periodic switching between two target patterns $\boldsymbol{\tau}_{1}$ and $\boldsymbol{\tau}_{2}$ every $t_{\mathrm{interval}}$ generations, and analyze the phenotypic and fitness landscapes of the evolved genotypes. As initial conditions, we use genotypes sampled after $10^4$ generations of evolution under the fixed target pattern $\boldsymbol{\tau}_1$. We simulate evolution for $10^4$ generations under periodic environmental switching using $T_{p}=1.5, \boldsymbol{\tau}_{1}=\left(+1, +1, +1\right), \boldsymbol{\tau}_{2}=\left(+1, +1, -1\right)$. It is expected that as environmental changes become more frequent, that is, as $t_{\mathrm{interval}}$ becomes shorter, the phenotypic landscape becomes more rugged, corresponding to lower penetrance of the dominant phenotypes.

Fig.~\ref{fig:Repeating}a and b show representative evolutionary trajectories for $t_{\mathrm{interval}}=100$ and $10$. In Fig.~\ref{fig:Repeating}b, at each environmental switch, the fitness of the current genotype drops abruptly because the target changes while the genotype is initially unchanged. Fitness then recovers over subsequent generations under the mutation-selection dynamics, directly visualizing mutation-mediated adaptation after each environmental change. We observe that when the interval between environmental changes is long, fitness in the current environment can nearly reach its maximum before the next change; when the interval between environmental changes is short, fitness does not have enough time to approach its maximum before the next environmental change. Fig.~\ref{fig:Repeating}c suggests that this trend is general. 

Fig.~\ref{fig:Repeating}d and \ref{fig:Repeating}e show the phenotypic and fitness landscapes of genotypes after evolution under periodic environmental switching. These results indicate that as $t_{\mathrm{interval}}$ decreases, the phenotypic landscape becomes more rugged, corresponding to lower penetrance, and the fitness landscape also becomes more rugged. 

Since the phenotypic landscapes of peripheral genotypes in $G\left(\boldsymbol{\tau}_{1}\right)$ are rugged as shown in Fig.~\ref{fig:Fixed}, it is then expected that as the frequency of environmental changes increases, peripheral genotypes are more likely to be selected. To test this, we define $\boldsymbol{\tau}_{\mathrm{preferred}}\left(\boldsymbol{J}\right)$, when it exists, as the unique target pattern $\boldsymbol{\tau}$ such that $\langle{F}\rangle\left(\boldsymbol{J}; \boldsymbol{\tau}, T_{p}\right)>0.5$ for $\boldsymbol{J}$ (see Materials and Methods), and measure the fraction of trajectories ending at a core genotype in $G\left(\boldsymbol{\tau}_{\mathrm{preferred}}\left(\boldsymbol{J}\right)\right)$ (Fig.~\ref{fig:Repeating}c) \footnote{Trajectories ending in $\mathcal{J}_{\mathrm{rem}}$, for which $\boldsymbol{\tau}_{\mathrm{preferred}}\left(\boldsymbol{J}\right)$ is undefined, were counted as non-core.}. This result is consistent with this expectation. 

When environmental changes are infrequent, that is, the environment is almost fixed, genotypes with high centrality are selected. Such genotypes have high penetrance and high mutational robustness. This is consistent with the relationship between centrality and fitness under a fixed environment in Fig.~\ref{fig:Fixed}. In contrast, as the frequency of environmental changes increases, genotypes with lower centrality, whose penetrance and mutational robustness are lower, are selected, thereby maintaining mutational accessibility to environmental change. 

We next examine whether there is a trade-off between penetrance and mutational accessibility, and whether the timescale of environmental changes affects penetrance and mutational accessibility along that trade-off during evolution. To reveal this, we calculate the mean penetrance of a dominant phenotype and $D$ for each $t_{\mathrm{interval}}$ shown in Fig.~\ref{fig:Repeating}f. This directly indicates that there is a trade-off between penetrance and mutational adaptability, and infrequent environmental changes favor penetrance, whereas frequent environmental changes favor mutational adaptability in that trade-off.

\begin{figure}[htp]
\centering
\includegraphics[width=.99\linewidth]{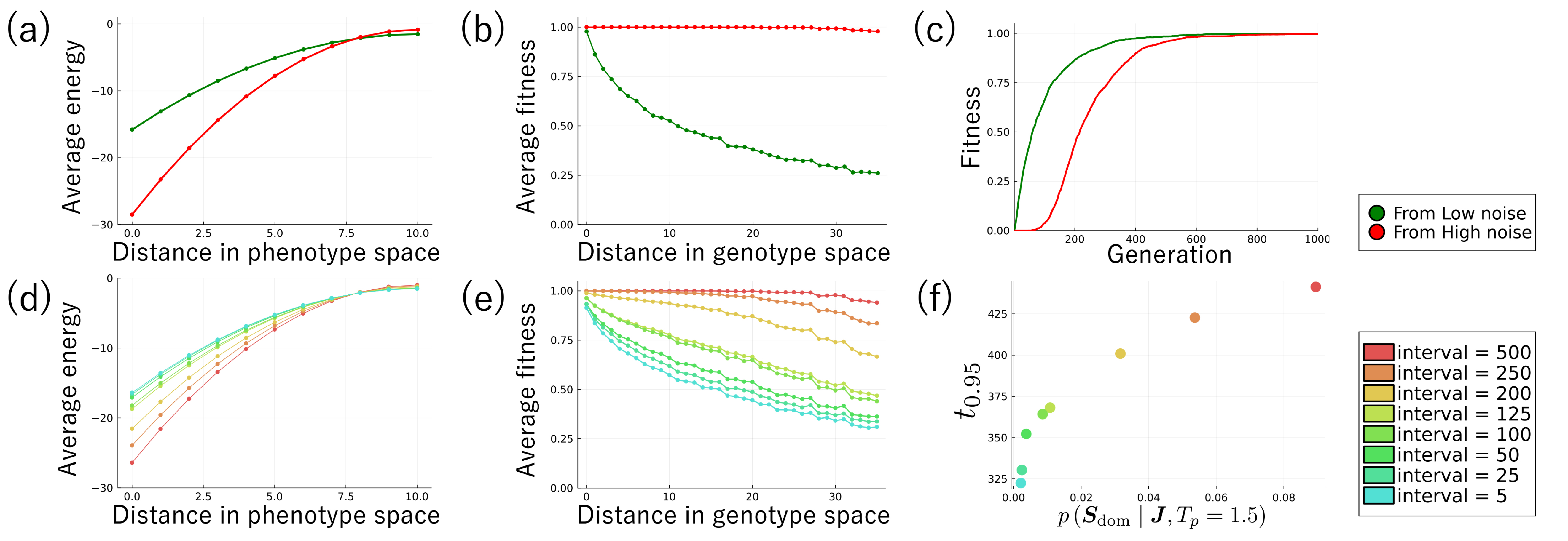}
\caption{(a) Average energy profile of phenotypic landscapes over $1000$ genotypes sampled after $10^4$ generations of evolution under low- and high-noise conditions in the same manner as Fig.~\ref{fig:Fixed}e, and (b) Average local fitness profile over $1000$ genotypes sampled after $10^4$ generations of evolution under low- and high-noise conditions in the same manner as Fig.~\ref{fig:Fixed}g. We use the average mutant fitness at each mutational distance $d$ over $100$ samples instead of the average over all possible mutants. (c) Average fitness trajectory in evolution after a single environmental change from $\boldsymbol{\tau}_{1}$ to $\boldsymbol{\tau}_{2}$ over $1000$ trajectories of genotypes evolved in high- and low-noise conditions. (d) Average phenotypic energy profile over $1000$ genotypes after $5 \times 10^3$ generations as in Fig.~\ref{fig:Repeating}d. (e) Average fitness profile at $T_{p}=0.125$ over $1000$ genotypes after $5 \times 10^3$ generations as in Fig.~\ref{fig:Repeating}e. (f) Relationship between penetrance and adaptability, quantified by the average penetrance of a dominant phenotype and $t_{0.95}$ for each $t_{\mathrm{interval}}$. In (f), we quantify $t_{0.95}$ by the number of generations required for fitness to reach $0.95 f$, where $f$ is the maximum over the fitness trajectory in each evolutionary trajectory. }
\label{fig:System_size}
\end{figure}

\FloatBarrier

\subsection*{Validation in a larger system}

We next examine whether the trends observed for $N=6, N_{\mathrm{focal}}=3$ persist for a larger system with $N=20$ and $N_{\mathrm{focal}}=4$. Because comprehensive enumeration is computationally infeasible for this larger system, we test robustness to system size through evolutionary simulations at selected values of the phenotypic noise level $T_{p}$. Although exhaustive enumeration of genotype space is infeasible for $N=20$, we still enumerate all $2^{20}$ phenotypes for each sampled genotype.

First, we test selection under a fixed environment. We simulate evolution in a given fixed target pattern $\boldsymbol{\tau}_{1}=\left(+1, +1, +1, +1\right)$ for $10^4$ generations with $T_{p}=0.03125$ (low-noise condition) and $T_{p}=1.5$ (high-noise condition). We test whether evolution in a fixed environment with high phenotypic noise again selects genotypes with funnel-like phenotypic landscapes and plateau-like fitness landscapes as in Fig.~\ref{fig:Fixed}e, g, h, i. Consistent with the $N=6$ results, evolved genotypes in the high-noise condition exhibited more funnel-like phenotypic landscapes and more plateau-like local fitness landscapes than those evolved in the low-noise condition (Fig.~\ref{fig:System_size}a, b).

Second, we examine adaptation after a single environmental change. To quantify mutational adaptability to a single environmental change, we simulate evolutionary adaptation to the new target pattern $\boldsymbol{\tau}_{2}=\left(+1, +1, -1, -1\right)$ at $T_p=0.125$. As initial genotypes, we use genotypes that had evolved in a given target pattern $\boldsymbol{\tau}_1$ for $10^4$ generations under either low- or high-noise conditions. Based on the $N=6$ results (Fig.~\ref{fig:Fixed}), genotypes evolved under high-noise conditions are expected to be located near the center of $G\left(\boldsymbol{\tau}_{1}\right)$. Such genotypes should adapt more slowly to the new target pattern $\boldsymbol{\tau}_{2}$. Fig.~\ref{fig:System_size}c shows that genotypes evolved under the high-noise condition tend to adapt to the new environment more slowly than those evolved under the low-noise condition.  

Finally, we simulate evolution under periodic environmental switching. As in the $N=6$ system, we analyze the phenotypic and fitness landscapes of genotypes evolved under periodic environmental switching in which two target patterns $\boldsymbol{\tau}_{1}$ and $\boldsymbol{\tau}_{2}$ alternate every $t_{\mathrm{interval}}$ generations. We simulate such evolution for $5 \times 10^3$ generations at $T_{p}=1.5$. The initial genotypes are sampled after evolution under a given fixed target pattern $\boldsymbol{\tau}_{1}$ for $10^4$ generations at $T_{p}=1.5$, as in Fig.~\ref{fig:Repeating}. We sample $1000$ evolutionary trajectories. These results indicate that as the frequency of environmental changes increases, the phenotypic and fitness landscapes become more rugged, as shown in Fig.~\ref{fig:System_size}d and e. In addition, we demonstrate the trade-off between penetrance and mutational adaptability by calculating the average penetrance of a dominant phenotype and $t_{0.95}$ for each $t_{\mathrm{interval}}$, shown in Fig.~\ref{fig:System_size}f. 

Overall, the results for $N=20$ are consistent with those for $N=6$, suggesting that the trends observed here are robust with respect to system size.

\section*{Discussion}

In this paper, through a comprehensive analysis of a statistical-physics genotype-phenotype (GP) map model, we demonstrate a trade-off between penetrance and mutational adaptability throughout genotype space: high-penetrance genotypes are located farther from high-fitness regions associated with alternative environments. This relationship is not obvious because penetrance is defined for an individual genotype, whereas mutational adaptability depends on where that genotype lies relative to high-fitness regions associated with alternative environments. In addition, the timescale of environmental change can therefore shift evolution toward opposite sides of the trade-off: a fixed environment can favor genotypes with high penetrance and high mutational robustness, whereas more frequent environmental change favors more adaptable genotypes with low penetrance. 

Our genotype-resolved analysis reveals an organization that was hidden in previous ensemble-averaged studies of the same spin model. Earlier studies used ensemble averages of energy and fitness to identify a regime in which evolved genotypes exhibit high phenotypic and mutational robustness \cite{Sakata2009-aq, Pham2023-iu}. Such ensemble averages identify the emergence of a robust regime but do not reveal how the individual genotypes within that regime differ from one another. Our results show that this heterogeneity is geometrically structured across genotype space: genotypes with high penetrance and mutational robustness occupy the core of the high-fitness region associated with the current environment, whereas genotypes with greater access to high-fitness regions associated with alternative environments are located toward the periphery and have lower penetrance. Thus, the present analysis resolves the internal geometry of the robust regime identified previously and connects it to mutational adaptability. 

At first glance, our finding that high-penetrance, mutationally robust genotypes have reduced mutational adaptability appears to conflict with the classical view that so-called neutral networks, which consist of neutral mutants, promote evolutionary innovation \cite{Wagner1996-gx, Ancel2000-on, Ciliberti2007-tp, Ciliberti2007-zm}. On a neutral network, evolution can explore many genotypes without a substantial loss of fitness, thereby gaining access to novel phenotypes. Our results suggest that this apparent conflict is resolved by considering phenotypic noise. When phenotypic noise is weak, genotypes adapted to the same target have nearly equal fitness throughout the high-fitness region, allowing nearly neutral exploration across the region. When phenotypic noise makes differences in penetrance selectively consequential, this effective neutrality is broken, and selection favors high-penetrance core genotypes over peripheral genotypes that provide greater mutational access to alternative adaptive regions. Thus, the two pictures describe complementary regimes rather than conflicting mechanisms. Neutral-network-mediated innovation should be more readily observed when phenotypic noise is weak, whereas the trade-off between penetrance and adaptability should become more apparent when phenotypic noise generates fitness differences among otherwise adapted genotypes.

Although our model is abstract, its mathematical structure is closely related to coarse-grained models used for concrete biological GP maps. Statistical-physics models of protein folding describe protein conformations as states in a sequence-dependent energy landscape and capture key features of folding, including glassy behavior, folding transitions, and folding funnels \cite{Go1975-pu, Taketomi1975-ij, Go1981-ey, Taketomi1988-if, Sasai1990-cn, Bryngelson1995-vu}. Likewise, noisy gene regulatory networks have also been described using coupled binary-spin models that reproduce distributions of gene-expression states \cite{Walczak2009-mv}. These models share the essential architecture of the present model: genotype-dependent interactions define an energy landscape and thereby a stochastic distribution over phenotypic states. The geometric mechanism identified here may therefore apply broadly to stochastic GP maps that admit such an interaction-based energy representation. Whether the same core-to-periphery organization persists under the specific constraints of protein sequence space and regulatory-network topology and dynamics is an important question for future work.

This geometric picture suggests a possible explanation for why incomplete penetrance persists despite the advantage of reliably expressing an adaptive phenotype. Our results provide such an explanation: genotypes with lower penetrance tend to have greater mutational access to alternative adaptive phenotypes and can therefore be favored when environments change. Previous studies have shown that genotypes with incomplete penetrance can serve as intermediates in the evolution of new traits, thereby facilitating phenotypic transitions during evolution \cite{Eldar2009-mc, Beye2013-zs}. These observations are consistent with our prediction that such genotypes occupy regions of genotype space from which phenotypes favored in other environments are more readily reached by mutation. More broadly, observing incomplete penetrance for a phenotype may suggest an evolutionary history in which that phenotype was favored only intermittently as environments changed.

Taken together, our results establish the global geometry of the GP map as a common basis for understanding the relation between penetrance and mutational adaptability. By bringing previously separate observations into a single framework, this geometric view opens a path toward identifying common organizing principles across empirical GP maps and developing a more predictive theory of evolution under environmental change.

\section*{Materials and Methods}

\subsection*{Definition of genotype network}

To characterize the position of each genotype in genotype space, we define the high-fitness region in a given target pattern as the network composed of genotypes with high fitness, and focus on this network. For each $\boldsymbol{\tau}_i$, we define $\mathcal{J}_{\boldsymbol{\tau}_{i}}$ as the set of genotypes satisfying $\langle{F}\rangle\left(\boldsymbol{J}; \boldsymbol{\tau}_{i}, T_{p}=0.03125\right)>0.5$, and let $G\left(\boldsymbol{\tau}_{i}\right)$ be the network where nodes are genotypes in $\mathcal{J}_{\boldsymbol{\tau}_{i}}$. Let $\mathcal{J}_{\mathrm{rem}}$ be the set of $\boldsymbol{J}$ such that $\langle{F}\rangle\left(\boldsymbol{J}; \boldsymbol{\tau}_{i}, T_{p}=0.03125\right) \leq 0.5\quad \text{for all } i$. Genotypes of $\mathcal{J}_{\mathrm{rem}}$ are not assigned to any high-fitness target region. Then, the genotype space is partitioned into the sets $\mathcal{J}_{\boldsymbol{\tau}_{i}}$ and $\mathcal{J}_{\mathrm{rem}}$, because
\begin{align*}
\sum_{i=1}^{2^{N_{\mathrm{focal}}-1}}\langle{F}\rangle\left(\boldsymbol{J}; \boldsymbol{\tau}_{i}, T_{p}\right)
&=\sum_{i=1}^{2^{N_{\mathrm{focal}}-1}}P\left(\boldsymbol{S}_{\mathrm{focal}}=\pm\boldsymbol{\tau}_{i}\mid\boldsymbol{J},T_{p}\right)\\
&=1 \left(\because Eq.~\ref{eqn:eq_fitness_J}\right). \numberthis \label{eqn:eq_sum_probability}
\end{align*}
Consequently, no genotype $\boldsymbol{J}$ satisfies both $\langle{F}\rangle\left(\boldsymbol{J}; \boldsymbol{\tau}_{i}, T_{p}\right)>0.5$ and $\langle{F}\rangle\left(\boldsymbol{J}; \boldsymbol{\tau}_{j}, T_{p}\right)>0.5$ for $i \neq j$. Hence, therefore it follows that if $i\neq j$, $\mathcal{J}_{\boldsymbol{\tau}_{i}}$ and $\mathcal{J}_{\boldsymbol{\tau}_{j}}$ are mutually exclusive. 

To focus on the regions in genotype space adapted to target patterns $\boldsymbol{\tau}_{1}$ and $\boldsymbol{\tau}_{2}$, we further define $G\left(\boldsymbol{\tau}_{1},\boldsymbol{\tau}_{2}\right)$ as the network induced by $\mathcal{J}_{\boldsymbol{\tau}_{1}}\sqcup\mathcal{J}_{\boldsymbol{\tau}_{2}}\sqcup\mathcal{J}_{\mathrm{rem}}$, thereby excluding genotypes predominantly adapted to the other target patterns.

\subsection*{Evolutionary simulation}

We simulate evolutionary dynamics as a Markov process in genotype space, where mutations are accepted with a probability determined by the resulting change in genotypic fitness. We consider a mutation-selection process in which at each generation, one of the possible mutations is chosen uniformly at random, and we define the acceptance probability of a mutant $\boldsymbol{J}'$ as 
\begin{align*}
w\left(\boldsymbol{J}\rightarrow\boldsymbol{J}'\right)=\frac{1}{\exp\left(-\frac{1}{T_{g}}\Delta\langle{F}\rangle\right)+1}, \numberthis \label{eqn:eq_transition}
\end{align*}
where $\Delta\langle{F}\rangle=\langle{F}\rangle\left(\boldsymbol{J}'; \boldsymbol{\tau}, T_{p}\right)-\langle{F}\rangle\left(\boldsymbol{J}; \boldsymbol{\tau}, T_{p}\right)$. Thus, mutations that increase genotypic fitness are accepted with higher probability. For a fixed target pattern $\boldsymbol{\tau}$, the equilibrium distribution of this Markov process is
\begin{align*}
P\left(\boldsymbol{J}\mid\boldsymbol{\tau},T_{g},T_{p}\right)&=\frac{1}{Z_{J}}\exp\left(\frac{1}{T_{g}}\langle{F}\rangle\left(\boldsymbol{J}; \boldsymbol{\tau}, T_{p}\right)\right), \numberthis \label{eqn:eq_eq_J} \\
Z_{J}&=\sum_{\boldsymbol{J}}\exp\left(\frac{1}{T_{g}}\langle{F}\rangle\left(\boldsymbol{J}; \boldsymbol{\tau}, T_{p}\right)\right).
\end{align*} 
In this evolutionary simulation, $T_{g}$ controls the strength of selection: smaller $T_g$ corresponds to stronger selection. In this study, we set $T_g$ to $0.01$. To compute $w\left(\boldsymbol{J}\rightarrow\boldsymbol{J}'\right)$, we enumerate all possible phenotypes, calculate their energies, and evaluate $p\left(\boldsymbol{S}\mid\boldsymbol{J}, T_{p}\right)$.

\subsection*{Uniqueness of $\boldsymbol{\tau}_{\mathrm{preferred}}\left(\boldsymbol{J}\right)$}

Because of Eq.~\ref{eqn:eq_sum_probability}, no genotype $\boldsymbol{J}$ satisfies both $\langle{F}\rangle\left(\boldsymbol{J}; \boldsymbol{\tau}_{i}, T_{p}\right)>0.5$ and $\langle{F}\rangle\left(\boldsymbol{J}; \boldsymbol{\tau}_{j}, T_{p}\right)>0.5$ for $i \neq j$. Therefore, whenever such a target exists, $\boldsymbol{\tau}_{\mathrm{preferred}}\left(\boldsymbol{J}\right)$ is unique.

\subsection*{Definition of $d_{\mathrm{half}}$}

We use the half-fitness mutational distance $d_{\mathrm{half}}$ to quantify mutational robustness in the small system. Let $\bar{F}\left(d\right)$ denote the average of $\langle{F}\rangle\left(\boldsymbol{J}_{\mathrm{mut}}; \boldsymbol{\tau}_{1}, T_p=0.03125\right)$ over all possible mutant genotypes $\boldsymbol{J}_{\mathrm{mut}}$s at mutational distance $d$ from the original genotype $\boldsymbol{J}$. Let $d_{0}$ be the smallest integer satisfying
\begin{align*}
\bar{F}\left(d_{0}\right) > \frac{\bar{F}\left(0\right)}{2}, \qquad \bar{F}\left(d_{0}+1\right) \leq \frac{\bar{F}\left(0\right)}{2}. \numberthis \label{eqn:def_dhalf_integer}
\end{align*}
We define $d_{\mathrm{half}}$ by linear interpolation between $d_{0}$ and $d_{0}+1$ as
\begin{align*}
d_{\mathrm{half}}=d_{0}+\frac{\bar{F}\left(d_{0}\right)-\frac{\bar{F}\left(0\right)}{2}}{\bar{F}\left(d_{0}\right)-\bar{F}\left(d_{0}+1\right)}. \numberthis \label{eqn:def_dhalf}
\end{align*}
Thus, $d_{\mathrm{half}}$ represents the mutational distance at which the interpolated average mutant fitness first decreases to half of the fitness of the original genotype.

\subsection*{Drawing genotype networks and landscapes}

We use the Fruchterman-Reingold force-directed algorithm to draw the genotype network $G\left(\boldsymbol{\tau}_{1}, \boldsymbol{\tau}_{2}\right)$ in Fig.~\ref{fig:Fixed}c and \ref{fig:Adaptability}a and \ref{fig:Repeating}a, as well as the two-dimensional phenotype network in Fig.~\ref{fig:Fixed}a.

In Fig.~\ref{fig:Fixed}b, nodes were arranged on concentric circles according to their distance from the original genotype. The original genotype was placed at the center, and nodes at $d=1$ were placed on the first circle, and nodes at distance 2 were placed on the second circle. The angular positions of nodes at $d=2$ were determined from the angular positions of their adjacent nodes at $d=1$, allowing the layout to reflect the local connectivity of the binary sequence network.

\section*{Author Contributions}

Y.I., K.K., and T.S.H. designed the research. Y.I. performed the simulations. Y.I. analyzed the data. Y.I., K.K., and T.S.H. wrote the paper.

\section*{Competing Interests}

The authors declare no competing interest.

\section*{Acknowledgments}

We thank Jumpei F. Yamagishi, Ryosuke Nishide, Yusuke Himeoka, and Chikara Furusawa for their helpful comments. This work was supported by the RIKEN Junior Research Associate Program (to Y.I.), the ANRI Fellowship (to Y.I.), the Japan Society for the Promotion of Science (JSPS) KAKENHI Grant Numbers 26K00057 and 26K00061 (to T.S.H.), and the Novo Nordisk Foundation Grant NNF21OC0065542 (to K.K.).

\FloatBarrier

\section*{Supporting Information}

\setcounter{figure}{0}
\renewcommand{\thefigure}{S\arabic{figure}}

\begin{figure}[htp]
\centering
\includegraphics[width=.45\linewidth]{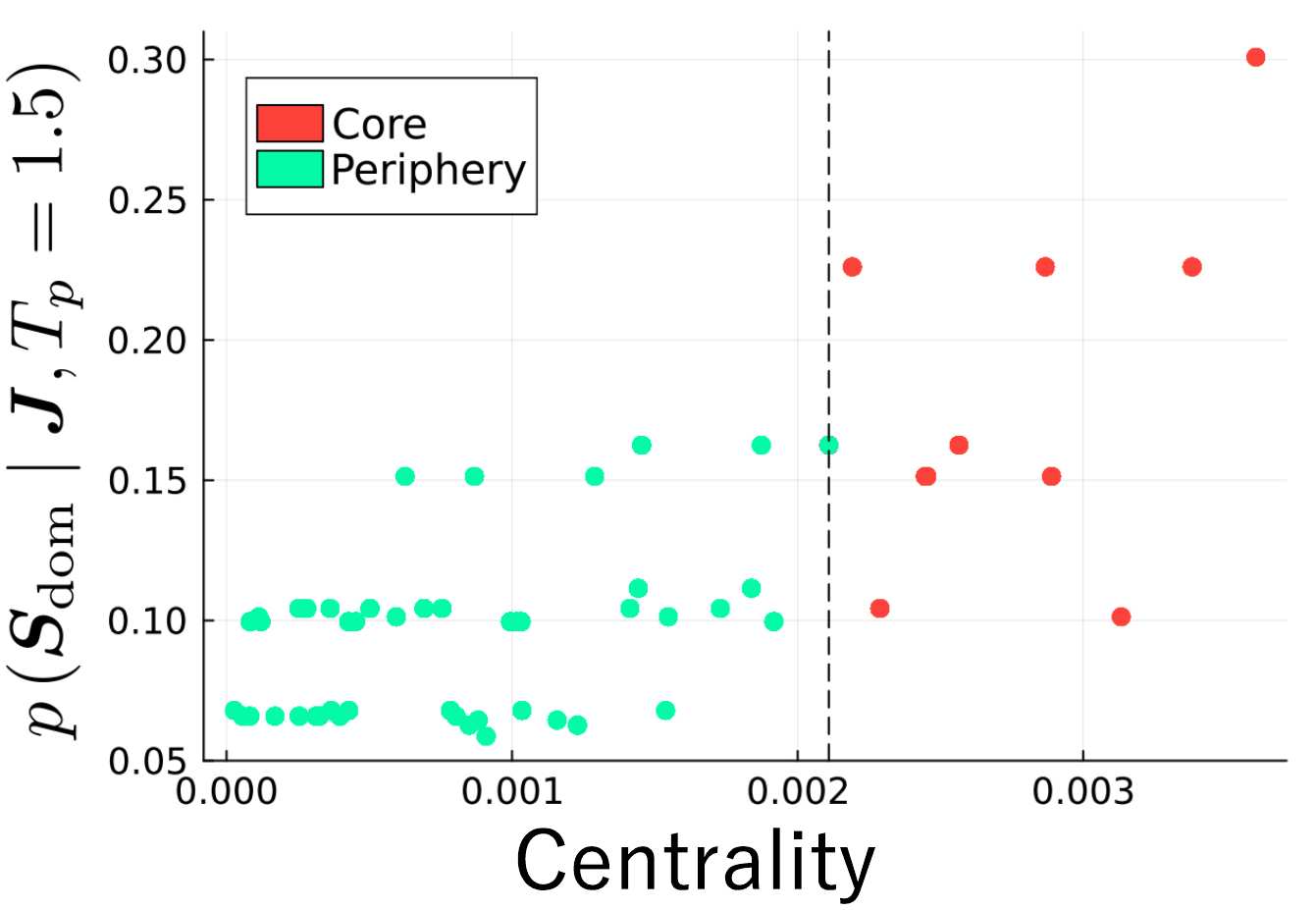}
\caption{The relationship between centrality and penetrance of a dominant phenotype $\boldsymbol{S}_{\mathrm{dom}}$ in $G\left(\boldsymbol{\tau}_{1}\right)$. Genotype centrality and penetrance of a dominant phenotype are correlated.}
\end{figure}

\begin{figure}[htp]
\centering
\includegraphics[width=.45\linewidth]{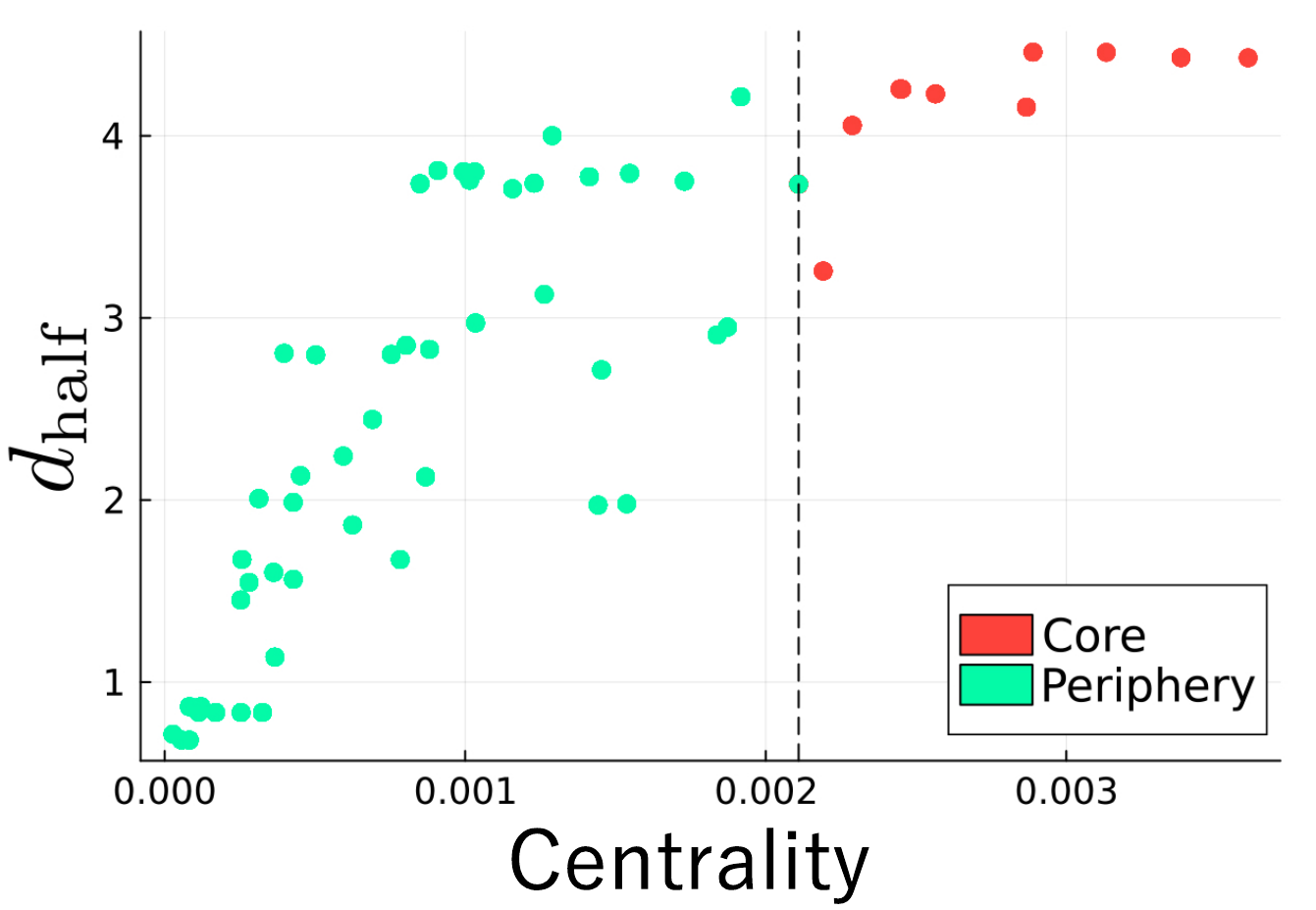}
\caption{The relationship between centrality and half-fitness mutational distance $d_{\mathrm{half}}$ in $G\left(\boldsymbol{\tau}_{1}\right)$. Genotype centrality and half-fitness mutational distance are correlated.}
\end{figure}

\begin{figure}[htp]
\centering
\includegraphics[width=.45\linewidth]{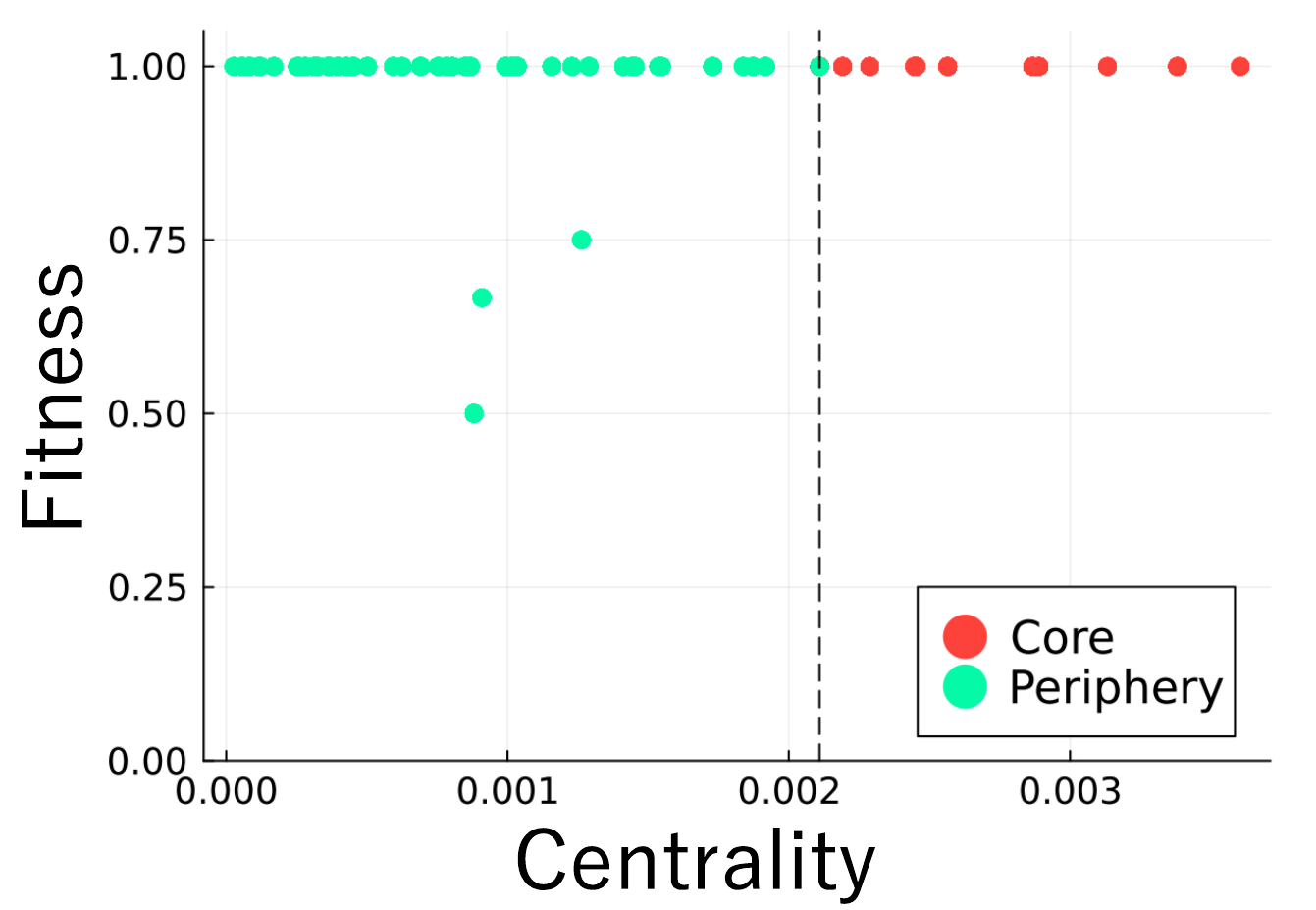}
\caption{The relationship between centrality and fitness in low-noise conditions.}
\end{figure}

\begin{figure}[htp]
\centering
\includegraphics[width=.80\linewidth]{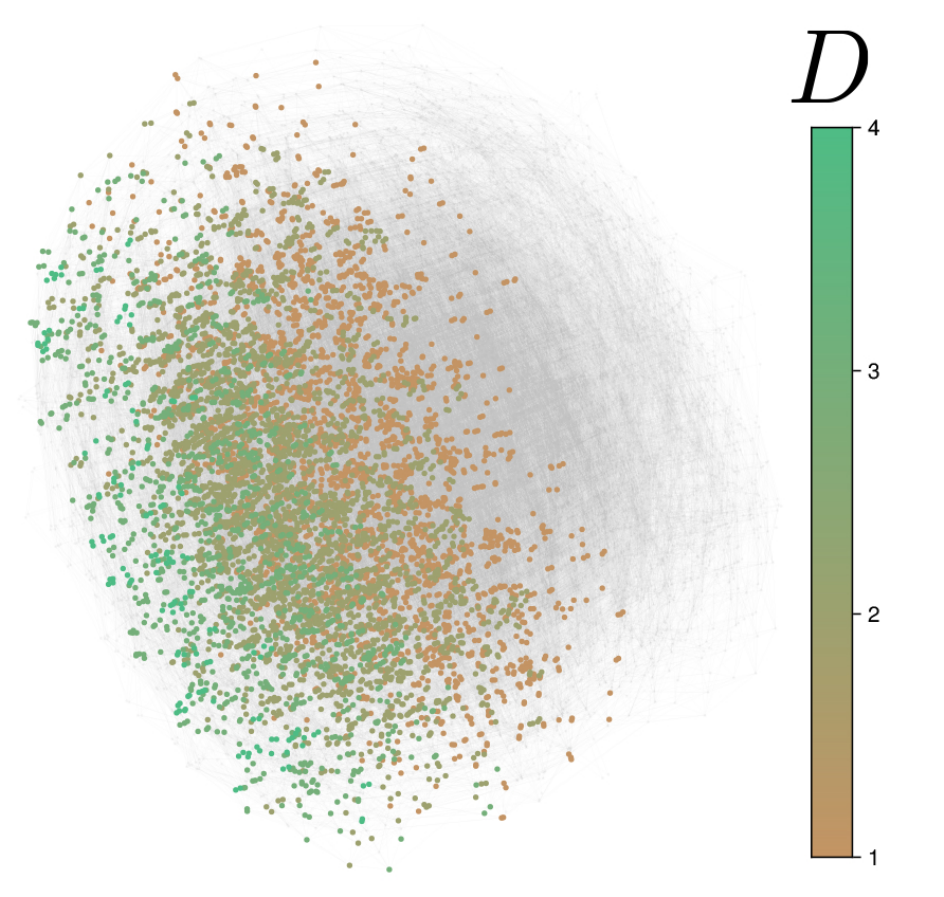}
\caption{Fitness in $D$ on the network $G\left(\boldsymbol{\tau}_{1},\boldsymbol{\tau}_{2}\right)$ in the same manner as Fig. 1c. }
\end{figure}

\FloatBarrier

\bibliographystyle{unsrt}
\bibliography{ref}

\end{document}